\documentstyle[psfig]{mn}

\newcommand{\markcite}[1]{}

\begin{document}

\title[Constraints from the cluster correlation function]
{Cosmological constraints from the correlation function of galaxy clusters}

\author[J. Robinson] {James
Robinson\thanks{jhr@palan.berkeley.edu}\\ Department of Astronomy,
University of California, Berkeley CA 94720}

\maketitle

\begin{abstract}
I compare various semi-analytic models for the bias of dark matter
halos with halo clustering properties observed in recent numerical
simulations. The best fitting model is one based on the collapse of
ellipsoidal perturbations proposed by Sheth, Mo \& Tormen (1999),
which fits the halo correlation length to better than 8 per cent
accuracy. Using this model, I confirm that the correlation length of
clusters of a given separation depends primarily on the shape and
amplitude of mass fluctuations in the universe, and is almost
independent of other cosmological parameters. Current observational
uncertainties make it difficult to draw robust conclusions, but for
illustrative purposes I discuss the constraints on the mass power
spectrum which are implied by recent analyses of the APM cluster
sample. I also discuss the prospects for improving these constraints
using future surveys such as the Sloan Digital Sky Survey. Finally, I
show how these constraints can be combined with observations of the
cluster number abundance to place strong limits on the matter density
of the universe.
\end{abstract}


\section{Introduction}

Recent years have seen dramatic increases in the size of the largest
cosmological N-body simulations, with giga-particle runs now able to
track the non-linear gravitational evolution of a significant fraction
of the observable universe (Colberg et al.\markcite{C98}
1998). Correspondingly, it is becoming possible to make more and more
detailed predictions for the non-linear properties of dark matter in
any cosmological model, with convergent results being obtained for
such features as the profiles and correlation functions of dark matter
halos (Moore et al.\markcite{M99} 1999). In spite of these advances,
the one great challenge remaining for the simulators is to
realistically track the evolution of cosmic gas through its complex
cycles of star-formation and feedback.

Given the current state of the art for cosmological simulations, rich
clusters of galaxies provide one of the most powerful tools with which
to make detailed tests of cosmology. Firstly, cluster potential wells
are so deep that their formation and evolution is largely unaffected
by the uncertain history of the cosmic gas, making dissipationless
N-body simulations an ideal place in which to study their
properties. Second, clusters correspond to rare peaks in the
primordial density field, so their statistics are extremely sensitive
to detailed features of the cosmology.

In this paper, I will show that a simple model proposed by Sheth, Mo
\& Tormen\markcite{SMT} (1999) can give an extremely good fit to the
correlation function of massive halos observed in recent state of the
art numerical simulations. Using this model I will compare the
predictions of a range of cosmological models with observations of the
cluster correlation length. In particular, I will show that
correlation length observations place strong constraints on the power
spectrum of mass fluctuations in the universe, constraints that are
almost independent of other cosmological parameters. In
section~\ref{sec-model} I describe the Sheth, Mo \& Tormen model for
the cluster correlation length, together with a similar semi-analytic
model due to Mo \& White (1996), and other bias models discussed in
the literature. In section~\ref{sec-simulations} I compare the
predictions of these models with recent numerical simulations, and in
section~\ref{sec-data} I show the constraints which can be derived
when the models are compared with observations. In
section~\ref{sec-conclusions} I discuss the relationship of this work
to previous studies, and draw my conclusions.

\section{Models for the correlation length}
\label{sec-model}
I describe four semi-analytic models which make predictions for the
amplitude of halo correlations, one due to Mo \& White\markcite{MW}
(1996 -- hereafter MW), one due to Sheth, Mo \& Tormen\markcite{SMT}
(1999 -- hereafter SMT), one due to Sheth and Tormen\markcite{ST}
(1999 -- hereafter ST), and one based on a model by Lee \&
Shandarin\markcite{LS} (1998 -- hereafter LS) and extended in ST. The
input parameters for these models are the power spectrum of matter
fluctuations $P(k)$ (extrapolated using linear gravity to today) and
the background cosmology of the universe (specified here in terms of
$\Omega_{\rm m}$ and $\Omega_\Lambda$, the contributions of matter and
a cosmological constant to the critical density today). I will compare
the predictions of these models with observations of the halo
correlation length $r_0(d)$, defined such that $\xi(r_0)=1$ where
$\xi(r)$ is the halo correlation function. Also, $d$ is the mean halo
separation for the halo sample in question, and the sample is chosen
to contain only those halos with ``richness'' greater than some
value. In this context I use ``richness'' (${\cal R}$) to denote any
parameter which correlates with halo mass (typical examples for
clusters are galaxy number counts and X-ray luminosity). I will assume
that the parameter $R$ can be monotonically transformed into an
``inferred mass'' ${\cal M}=f({\cal R})$ which is correlated with the
true mass with known conditional probability $p({\cal M}|M)$.

For each model, the procedure for computing the correlation length for
a sample of halos with mean separation $d$ is as follows.
\begin{itemize}
\item
First, compute the inferred mass threshold ${\cal M}$, such that the
mean separation of halos with inferred mass greater than ${\cal M}$ is
equal to $d$. That is, find ${\cal M}$ satisfying
\begin{equation}
N^{\rm I}_X({\cal M})=d^{-3}
\end{equation}
where $N^{\rm I}_{\rm X}({\cal M})$ is the number density of halos
with inferred (as denoted by the superscript I) mass greater than
${\cal M}$ for model X, and X denotes the semi-analytic formalism
being used (either MW, SMT, ST or LS). Now,
\begin{equation}
N^{\rm I}_{\rm X} ({\cal M}) = \int_{\cal M}^\infty n^{\rm I}_X({\cal
M}) d{\cal M}
\end{equation}
where $n^{\rm I}_X({\cal M})d{\cal M}$ is the number density of halos with
inferred masses between ${\cal M}$ and ${\cal M}+d{\cal M}$, and
\begin{equation}
n^{\rm I}_X({\cal M})=\int dM p({\cal M}|M) n_X(M)
\end{equation}
with 
\begin{equation}
n_X(M)=-\nu f_X(\nu) \frac{\bar{\rho}}{M}\frac{d\sigma(M)/dM}{\sigma(M)}
\end{equation}
being the number density of halos with true mass between $M$ and
$M+dM$.  Here $\nu=\delta_c/\sigma(M)$, $\sigma(M)$ is the {\it rms}
linear fluctuation in a sphere containing an average mass $M$,
$\bar{\rho}$ is the background density of the universe, and $\delta_c$
is the critical overdensity for collapse as computed in the spherical
collapse model -- see Kitayama\markcite{KS} \& Suto (1996) for fits to
the weak cosmological dependence. The quantity $\sigma(M)$ can be
straightforwardly calculated at any redshift knowing $P(k)$ and the
background cosmology.  The function $f_X$ is given by
\begin{equation}
f_{\rm MW} (\nu) = \sqrt{\frac{2}{\pi}}\exp\left(-\frac{\nu^2}{2}\right)
\end{equation}
for the MW model, by
\begin{equation}
f_{\rm SMT} (\nu) = f_{\rm ST} (\nu) =
A\left(1+\frac{1}{a^q\nu^{2q}}\right)\sqrt{\frac{2}{\pi}}\exp\left(- \frac{a\nu^2}{2}\right)
\end{equation}
for the SMT and ST models, with $A\simeq 0.3222$, $q=0.3$, and 
$a=0.707$, and by equation A1 of ST for the LS model.
\item
Calculate the effective bias parameter $b^X_{>{\cal M}}$ for halos with an
inferred mass greater than ${\cal M}$, satisfying
\begin{equation}
b^X_{>{\cal M}}=\frac{1}{N^I_X({\cal M})}\int_{\cal M}^\infty b^I_X({\cal M}) n^I_X({\cal M}) dM
\end{equation}
where 
\begin{equation}
b^I_X({\cal M})=\frac{\int dM p({\cal M}|M) f_z(b_X(M)) b_X(M) n(M)}{\int dM p({\cal M}|M) n(M)}
\end{equation}
and $b_X(M)$ is the bias parameter for halos with true mass $M$, given
by
\begin{equation}
b_{\rm MW}=1+\frac{1}{\delta_c}(\nu^2 -1 ) 
\end{equation}
for the MW model, by
\begin{eqnarray}
b_{\rm SMT}=&1+\frac{1}{\sqrt{a}\delta_c} &\left[
a^{5/2}\nu^2 + \sqrt{a} b
(a\nu)^{2(1-c)}\right. \\ &&\left.-  \frac{
(a\nu)^{2c}}{
(a\nu)^{2c} + b (1-c) (1-c/2)}\right]
\end{eqnarray}
for the SMT\footnote{The functions $f_{\rm SMT}(\nu)$ and $b_{\rm SMT}
(\nu)$ given here are just $f(\nu')$ and $b_{\rm Eul}(\nu')$ from
equations 5 and 8 of SMT respectively, with $\nu'=\sqrt{a} \nu$. As
discussed in SMT, this rescaling ensures that the halo mass function
agrees with that observed in numerical simulations.} model with
$a=0.707$, $b=0.5$ and $c=0.6$, by
\begin{eqnarray}
b_{\rm ST}&=&1+ \frac{1}{\delta_c}\left[(a \nu^2 -1) + \frac{2 q} {1+
\left(a \nu^2\right) ^q}\right]
\end{eqnarray}
for the ST model with $a$ and $q$ given above, and by equation A3 of
ST for the LS model.  Lastly, $f_z$ is an optional factor which may be
included to account for the effect of redshift space distortion if the
prediction is to be compared with observations in redshift
space. Following Kaiser\markcite{K87} (1987) $f_z$ is given by
\begin{equation}
f_z(b)=1+2\beta/3+\beta^2/5
\end{equation}
where $\beta=\Omega_m^{0.6}/b$.
\item
Solve~for~the~correlation~length~$r_0$~satisfying~$(b^X_{>{\cal M}})^2
\xi_{\rm NL} (r_0)=0$, where $\xi_{\rm NL} (r)$ is the non-linear mass
correlation function, given by the Fourier transform of the non-linear
power spectrum $P_{\rm NL} (k_{\rm NL})$. The non linear power
spectrum is computed using the method described by Peacock \&
Dodds\markcite{PD96} (1996) using the linear power spectrum $P(k)$ as
input. For comparison, we also compute $r_0$ using the linear
correlation function, that is we solve for $r_0$ satisfying
$(b^X_{>{\cal M}})^2\xi_{\rm L} (r_0)=1$, where $\xi_{\rm L} (r)$ is
the Fourier transform of the linear power spectrum $P(k)$. As we will
see in section~\ref{sec-simulations}, use of the non-linear
correlation function gives a better fit for all the models. Unless
otherwise specified, all correlation lengths have been computed using
the non-linear correlation function.
\end{itemize} 
 
\section{Comparison with simulations}
\label{sec-simulations}
\begin{figure}
\centerline{\psfig{file=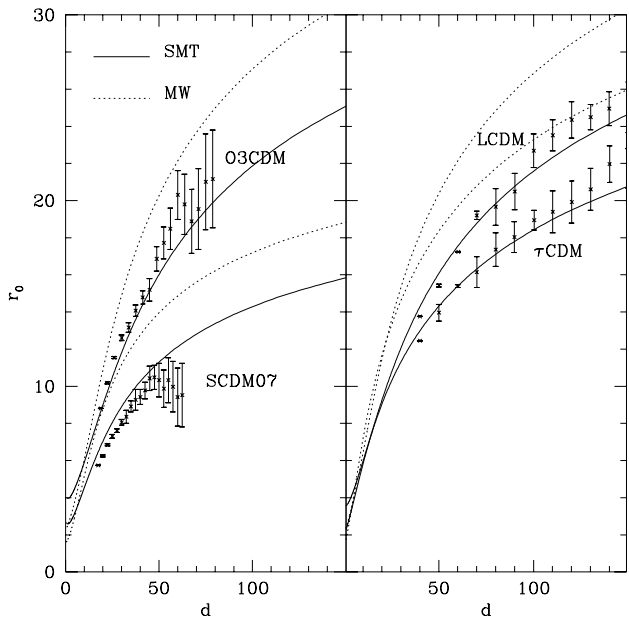,width=3.5in}}
\caption{Correlation length versus halo separation as observed in
numerical simulations (datapoints) together with the predictions of the SMT
(solid curves) and MW (dotted curves) models. The SMT predictions
agree very closely with the results of the simulations, while the MW
predictions are typically 25 per cent too high.}
\label{fig-simulation}
\end{figure}

The models just described make predictions for the halo two-point
correlation function on any length scale $r$. Since the shape of the
halo correlation function is not well constrained by current
observations, most authors have focused just on the amplitude,
quantified in terms of the correlation length $r_0$. I now compare the
predictions of each of the models for $r_0$ with the values observed
in large collisionless N-body simulations (comparison between models
and simulations for the full halo correlation function is beyond the
scope of the current paper, but deserves future investigation). The
simulations have been carried out by Colberg et al. (1998 -- hereafter
C98) and Governato et al. (1999 -- hereafter G99). For each simulation
the power spectrum has been taken to have a CDM form (Bardeen et
al.\markcite{BBKS}~1986, eq.~G3) parameterized by a shape parameter
$\Gamma$ (where $\Gamma
\simeq
\Omega h$, and $h$ is the Hubble constant in units of 100
kms$^{-1}$Mpc$^{-1}$) and normalization $\sigma_8$, where $\sigma_8$
is the {\it rms} fluctuation in an $8h^{-1}$Mpc sphere. The power
spectrum parameters for each simulation are summarized in
Table~\ref{tab-simulation}, together with the box length $L$, particle
number $N_p$ and background cosmology.  Fig.~\ref{fig-simulation}
compares the $r_0-d$ relation observed in the simulations (data
points) with that predicted by the MW (dotted lines) and SMT (solid
lines) formalisms. In computing these predictions, I use $p({\cal
M}|M)=\delta({\cal M}-M)$ where $\delta$ denotes the Dirac delta
function, since in this case the richness property by which the halos
are ranked is just the true mass.

Clearly the SMT model gives much
better agreement with the cluster correlation length as observed in
the simulations. I quantify the goodness of fit by defining an {\it
rms} error $E$ via
\begin{equation}
E^2=\frac{1}{N}\sum_{\rm i=1}^{N} \left(\frac{r^{\rm i}_0 - r^{\rm X}_0(d^i)}
{r^{\rm X}_0(d^i)}\right)^2
\end{equation}
where the sum $i$ runs over all datapoints ($d^i$,$r^i_0$) under
consideration, and $r^{\rm X}_0(d)$ denotes the correlation length
prediction for model X at separation $d$. For the combined datapoints
from all four simulations, the SMT model gives $E=0.08$, while errors
for the MW model are much larger, with $E=0.22$. The goodness of fit
for each of the models is given in Tab~\ref{tab-fits}. The left column
shows results using the non-linear correlation function to compute
$r_0$, while the right column shows results using the linear
correlation function. Use of the non-linear correlation function
improves the fit in each case.
\begin{table*}
\begin{minipage}{110mm}
\centering
\begin{tabular}{ccccccccccc}
\hline 
Model&Source&$\Omega_{\rm m}$& $\Omega_\Lambda$ & $\sigma_8$ & $\Gamma$ &
$N_p$ & $L/(h^{-1}\rm{Mpc})$\\
SCDM07    & G99 & 1.0 & 0.0 & 0.7 & 0.5 & $47 \times 10^6$ & $500$ \\
O3CDM     & G99 & 0.3 & 0.0 & 1.0 & 0.225 & $47 \times 10^6$ & $500$ \\
$\tau$CDM & C98 & 1.0 & 0.0 & 0.6 & 0.21  & $10^9$ & $2000$ \\
LCDM      & C98 & 0.3 & 0.7 & 0.9 & 0.21  & $10^9$ & $3000$ \\
\hline
\end{tabular}
\vspace{10pt}
\caption{Summary of parameters for each of the simulations discussed
in section~\ref{sec-simulations}.}
\label{tab-simulation}
\end{minipage}
\end{table*}

I conclude from this analysis that the SMT model gives a good fit
(typical errors less than 8 per cent) to the halo correlation length
arising from a full treatment of the non-linear gravitational
evolution. The MW model does worst of all the models considered, with
typical errors of order 25 per cent. In particular, the MW model
systematically overestimates the halo correlation length for fixed
separation $d$, a result also found in C98. It is worth noting that
the discrepancy between the MW model and the simulations is much
larger than might be inferred from previous studies. For instance, Mo,
Jing
\& White\markcite{MJW} (1996 - hereafter MJW)
and Jing\markcite{Jing} (1999) suggest
that the MW formula agrees at better than the five per cent level with
the rare halo bias observed in their simulations. Closer examination
of the rarest mass datapoints in Fig.~3 of Jing (1999) illustrate that
the error is actually much larger. Figs.~7 and 8 of Governato et
al. also imply extremely good agreement between the MW formula and
correlation lengths observed in their simulations. In fact, the MW
prediction has been computed incorrectly in these figures, and the
true agreement is much worse (as illustrated in
Fig.~\ref{fig-simulation} of this paper). A final source of confusion
as to the accuracy of the MW formula is that the curves showing the MW
prediction in Fig.~8 of MJW have also been
computed incorrectly, with the correct values for the correlation
length being up to 25 per cent higher in some places. To summarize,
there is clear evidence that the MW formula significantly
overestimates the expected bias of the rarest halos, while the SMT
model gives extremely good agreement.
\begin{table}
\centering
\begin{tabular}{ccc}
\hline 
Model& $E$--$(\xi_{\rm NL})$ & $E$-- $(\xi_{\rm L})$ \\
MW & 0.22 & 0.24  \\
SMT& 0.08 & 0.10 \\
ST & 0.11  & 0.12 \\
LS & 0.11  & 0.14\\
\hline
\end{tabular}
\vspace{10pt}
\caption{Goodness of fit for each of the bias models discussed in
section~\ref{sec-model}. The fit parameter $E$ is defined in
section~\ref{sec-simulations}. The first and second columns give the
fit obtained using the non-linear and linear correlation function
respectively.}
\label{tab-fits}
\end{table}

\section{Comparison with data}
\label{sec-data}
\begin{figure}
\centerline{\psfig{file=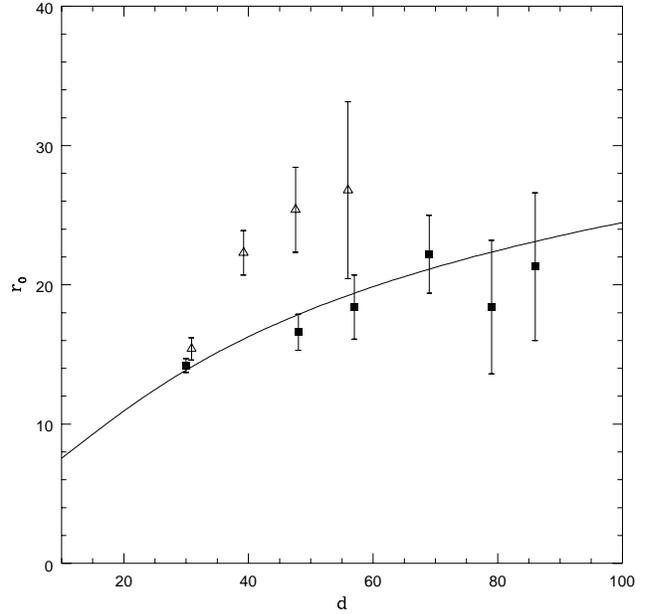,width=3.5in}}
\caption{Cluster correlation length versus halo separation as inferred
by Croft et al. (1997 - solid squares) and Lee and Park (1999 - open
triangles) from an analysis of the APM cluster sample. The solid line
shows the SMT prediction for a model with $\Omega_m=0.3$,
$\Gamma=0.17$ and $\sigma_8=1.0$.}
\label{fig-data}
\end{figure}

I now make use of the SMT model to compare the predictions of a range
of cosmologies with observational data. Fig.~\ref{fig-data} shows the
$r_0-d$ relation for the APM cluster survey, as analyzed by Croft et
al.\markcite{C97} (1997 - hereafter C97) and Lee \& Park\markcite{LP}
(1998 - hereafter LP98). Although this is only a small fraction of
currently available data on the cluster correlation length, it
suffices to illustrate the wide degree of systematic uncertainty which
exists in present observations -- even two analyses of the same
dataset give results that differ by considerable factors, and the
situation is even less clear when results from Abell (see for instance
Bahcall \& Cen 1992) and X-ray (see LP99 for a detailed discussion)
cluster samples are included. Although present day uncertainties are
large, errors in $r_0$ will be greatly reduced by future surveys. The
Sloan Digital Sky Survey (SDSS, see project book at
http://www.astro.princeton.edu/PBOOK/welcome.htm for detailed
specifications) is expected to identify roughly 1000 nearby clusters
in its spectroscopic galaxy redshift survey, and even more in its
photometric survey.  The inclusion of redshift information in the
cluster identification procedure will greatly reduce systematic
uncertainties in $r_0$, and statistical uncertainties will be reduced
by approximately 50 per cent by the increased sample size. In order to
illustrate the type of results we can hope for, I will carry out the
bulk of my analysis using just the C97 data. This data is at least
self-consistent, and 2$\sigma$ confidence limits for the SDSS survey
should be similar in size to 1$\sigma$ confidence limits from the C97
analysis.

I compute confidence limits for various models as follows: Each model
gives a specific prediction for the for the $r_0-d$ relation, given
which the likelihood\footnote{This analysis assumes that the
datapoints are statistically independent. While this should be a good
approximation for the datasets considered, the effect of correlations
should ultimately be taken into account in a more detailed treatment.}
of observing a particular dataset is (up to a constant of
proportionality)
\begin{equation}
\label{eqn-like}
{\cal L}\propto e^{-\chi^2/2}
\end{equation}
where 
\begin{equation}
\label{eqn-chi2}
\chi^2=\sum_{\rm i}\frac{\left(r^{\rm i}_0 - r^{\rm X}_0(d^i)\right)^2}
{\sigma^2_i+\left(E r^{\rm X}_0(d^i)\right)^2}.
\end{equation}
Here the sum $i$ runs over all datapoints ($d_i$,$r^i_0$) with
standard deviation $\sigma_i$, $E$ is the fractional uncertainty in
the predictions of the model (which following the discussion in
section~\ref{sec-simulations} is taken to be $E=0.08$) and the term
$\left(E r^{\rm X}_0(d^i)\right)^2$ is added in quadrature to the
denominator to account for systematic uncertainties in the
model. Given a set of model parameters and prior distributions for
those parameters, I calculate $1\sigma$ and $2\sigma$ confidence
limits by computing the likelihood threshold above which the
integrated likelihood accounts for 69 and 95 per cent respectively of
the total probability. Finally, all model predictions are computed at
the median redshift of each sample, which I take to be $z=0.08$,
although the precise value used has little effect on the results.

\begin{figure}
\centerline{\psfig{file=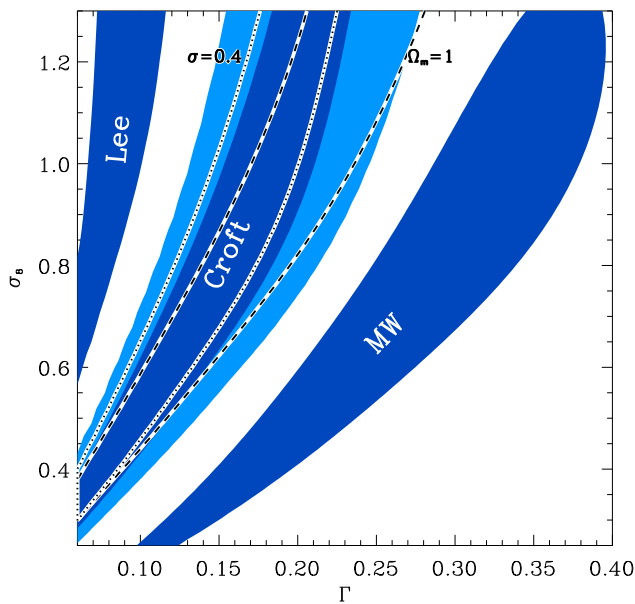,width=3.5in}}
\caption{Confidence limits in the $\Gamma-\sigma_8$ plane derived from
cluster correlation length data. The shaded area labeled ``Croft''
shows 1$\sigma$ (dark band) and 2$\sigma$ (light band) limits derived
from the fiducial analysis utilizing the Croft data and assuming the
SMT model for cluster correlations, in an $\Omega_m=0.3$,
$\Omega_\Lambda=0.7$ cosmology with zero scatter between true and
inferred cluster mass. Each of the other four bands show the
results of changing one factor in the fiducial analysis; using the Lee
\& Park (1999) analysis of the APM clusters
(shaded area labeled Lee), using the MW correlation length formula
(shaded area labeled MW), using a flat $\Omega_m=1.0$ cosmology (area
delineated by long-dashed lines), and finally using a 50 per cent
scatter between true and inferred cluster mass (area delineated by
short-dashed lines).}
\label{fig-square}
\end{figure}

Fig.~\ref{fig-square} shows confidence limits in the $\Gamma-\sigma_8$
plane derived from the observations shown in
Fig.~\ref{fig-data}. Given values for $\sigma_8$ and $\Gamma$, each
model is fully specified once we know $\Omega_{\rm m}$ and the
conditional probability distribution relating the ``inferred'' mass
for the cluster observations in question to the true cluster mass. The
shaded area labeled ``Croft'' shows 1$\sigma$ (dark region) and
2$\sigma$ (light region) confidence limits derived using the SMT model
and the C97 data, in an $\Omega_m=0.3$ background cosmology with zero
scatter between inferred and true cluster mass (use of an open
$\Omega_{\rm m}=0.3$ cosmology makes virtually no difference to the
results).
Four other confidence regions are also shown in this figure, each one
the result of changing one aspect of the first analysis. First, the
shaded area labeled ``MW'' gives the $1\sigma$ confidence region
which results from using the MW model rather than the SMT model. The
limits in this case are quite different, and given the poor
performance of the MW model when compared to simulations, should be
disregarded. Second, the shaded region labeled ``Lee'' shows the
1$\sigma$ limits resulting from using the LP99 data (with its higher
values for $r_0$) rather than the C98 data. The LP99 data favours
dramatically lower values of $\Gamma$ for a given value of $\sigma_8$,
and since the majority of alternative datasets (see LP99 for a
detailed discussion) also prefer higher values of $r_0$, the high
$\Gamma$ region to the right of the ``Croft'' confidence limits is
likely to be excluded by all current observations. Thirdly, the
long-dash lines delineate the 1$\sigma$ confidence region resulting
from an analysis identical to the ``Croft'' analysis except for the
choice of a flat $\Omega_{\rm m}=1$ cosmology. Increasing $\Omega_{\rm
m}$ slightly increases the amplification of clustering by redshift
space distortion, and slightly reduces the growth factor at the median
redshift of the sample. Both of these effects are small, and the
resulting confidence region is very similar to the $1\sigma$ limits
resulting from the $\Omega_{\rm m}=0.3$ analysis. Lastly, the short
dashed lines delineate 1$\sigma$ confidence regions obtained when a
significant scatter is introduced between the inferred and true
cluster masses. In particular, $p({\cal M}|M)$ is modeled as a
log-normal distribution with a natural logarithmic standard deviation
$\sigma=0.4$. This value of $\sigma$ corresponds to roughly a 50 per
cent scatter in the inferred mass for a given true mass. Even for such
a large scatter, the 1$\sigma$ confidence region is virtually
unchanged, demonstrating that robust constraints can be obtained even
if clusters of a given mass have a wide distribution of ``richness''
values - the only requirement is that there is some monotonic
transformation which loosely correlates the richness (for instance,
X-ray luminosity or galaxy counts) with the true mass.

\begin{figure}
\centerline{\psfig{file=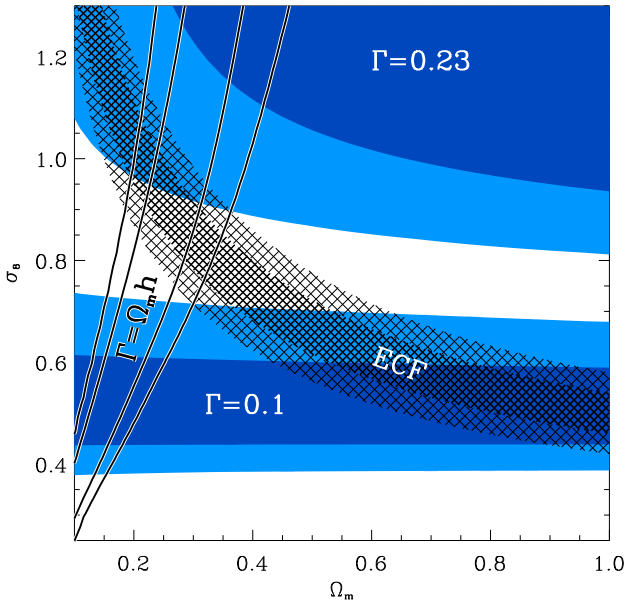,width=3.5in}}
\caption{Confidence limits in the $\Omega_{\rm m}-\sigma_8$ plane
derived from cluster correlation length data and cluster abundance
data. The hashed area labeled ``ECF'' shows 1$\sigma$ (dark band) and
2$\sigma$ (light band) limits from the cluster abundance analysis of
Eke, Cole \& Frenk\markcite{ECF} (1996). Each of the other three bands
shows correlation length constraints from the C97 data, under three
different assumptions about the value of $\Gamma$, as labeled. For the
$\Gamma=\Omega_{\rm m} h$ case I take h=0.65.}
\label{fig-square_omega}
\end{figure}

Fig.~\ref{fig-square} demonstrates that observations of the cluster
correlation length place constraints on the amplitude \& shape of the
matter power spectrum in the universe which are almost independent of
cosmology. As one final point, I illustrate the type of cosmological
constraints which can be obtained when these limits are combined with
independent observations of the mass power
spectrum. Fig.~\ref{fig-square_omega} shows confidence limits in the
$\Omega_{\rm m}-\sigma_8$ plane from a combination of cluster
correlation length data and cluster number abundance data. The dark
and light hashed regions show 1$\sigma$ and 2$\sigma$ confidence bands
derived from the local cluster temperature function by Eke, Cole \&
Frenk\markcite{ECF} (ECF - 1996). The other confidence bands show
1$\sigma$ and 2$\sigma$ limits derived from the C97 data, under three
different assumptions about the value of $\Gamma$\footnote{Current
uncertainties in $\Gamma$ should greatly be greatly reduced by future
surveys such as SDSS, allowing us to place tight constraints on
$\Omega_m$ using the techniques described here.}. The solid lines show
constraints for the case of a CDM cosmology with $\Gamma=\Omega_{\rm
m} h \exp(-\Omega_{\rm B}(1+\sqrt{2h}/\Omega_{\rm m}))$ with $h=0.65$,
and $\Omega_{\rm B}=0.04$ (for conciseness, this region is simply
labeled by ``$\Gamma=\Omega_{\rm m} h$''). Increases (decreases) in
the value of $h$ just shift the confidence region to lower (higher)
values of $\Omega_{\rm m}$ by the same factor. The shaded bands show
confidence limits for the choices $\Gamma=0.23$ (which provides a good
fit to the power spectrum of APM galaxies - see Viana \&
Liddle\markcite{VL} 1996) and $\Gamma=0.1$. For $\Gamma=0.1$ the
combined ECF and C97 constraints are consistent with the case
$\Omega_{\rm m}=1$, while the other choices for $\Gamma$ each require
$\Omega_{\rm m}<0.35$ for consistency between the datasets at the
2$\sigma$ level. If the LP99 data was used instead of C97, even lower
values of $\Omega_{\rm m}$ would be preferred.

\section{Conclusion}
\label{sec-conclusions}
I have shown that the halo correlation length measured in recent
numerical simulations can be well fit (to better than 8 per cent
accuracy) by a semi-analytic model based on the collapse of
ellipsoidal perturbations due to Sheth, Mo \& Tormen\markcite{SMT}
(1999). A similar analytic model due to Mo
\& White\markcite{MW} (1996) on the other hand gives a poorer fit to
the simulations (roughly 25 per cent accuracy). Applying the SMT
result to present data, I have shown that correlation length
observations place strong, almost cosmology independent constraints on
the shape and amplitude of the matter power spectrum in the
universe. By far the greatest source of uncertainty is systematic
discrepancies between current datasets, but if the cluster correlation
length is at least as high as is implied by the APM survey, then an
interesting region of power spectrum parameter space (everything to
the right of the ``Croft'' region in Fig.~\ref{fig-square}) can be
excluded. Future surveys, including the Sloan Digital Sky Survey,
should greatly reduce the observational uncertainties, with 2$\sigma$
confidence limits from SDSS being as tight as the 1$\sigma$ limits
shown for the C97 data in Fig.~\ref{fig-square}. One particularly
interesting result is that the analysis is almost unaffected when a
significant scatter (up to 50 per cent) is introduced between true
cluster mass and the richness property by which clusters are
ranked. Such a scatter is inevitable in any cluster survey, so it is
extremely useful to know that even an effect this large does not
influence the results. Finally, I have shown how the correlation
length can be combined with other cluster observations to place limits
on the matter density of the universe.

A number of previous studies have discussed constraints from the
cluster correlation length. Bahcall \& Cen\markcite{BC} (1992) and
Croft \& Efstathiou\markcite{CE} (1994) carried out numerical
simulations showing that cluster correlation length was a strong
discriminator amoung models, and obtained results which are consistent
with those discussed in this paper. Mo, Jing \& White\markcite{MJW}
(1996) and later Robinson, Gawiser \& Silk (1998) made use of the MW
formalism to compare models and observations for a wider range of
parameters, and owing to the inaccuracy of the MW model, their results
are not consistent with those found here (the inconsistency of the MW
model is actually larger than is apparent in MJW due to an error in
Fig.~8 of that paper). The work described here improves on previous
studies by utilizing a formula which accurately fits the most recent
numerical simulations, and which can be used to compute predictions
for a large range of model parameters very quickly.
 
Lastly, it should be noted that the analysis discussed here assumes
that the primordial fluctuations in the universe are
Gaussian. Non-gaussianity also influences the value of the cluster
correlation length, and a number of authors, including Robinson,
Gawiser \& Silk (1998) and Koyama, Soda \& Taruya \markcite{KST}
(1999) have attempted to exploit this fact to use cluster observations
to place constraints on primordial non-gaussianity. Both these works
have modeled the correlation function for non-gaussian models using an
extension of the MW formalism, and therefore their conclusions should
be modified in the light the results discussed here, which show that
MW does not accurately predict the halo correlation function even for
Gaussian models.

To summarize, cluster correlation length observations place strong
cosmology independent constraints on the matter power spectrum in the
universe, constraints which future surveys such as SDSS will allow
us to fully exploit.

\section{acknowledgements}
I would like to thank Marc Davis, Eric Gawiser and Joseph Silk for
helpful and stimulating discussions. I would also like to thank Fabio
Governato and H.~J.~Mo for their swift answers to my questions, and
Pedro Ferreira for reading a draft of the manuscript. This work has
been supported in part by grants from the NSF, including grant
9617168.

\end{document}